\documentclass[conference]{IEEEtran}
\usepackage{graphicx}
\usepackage{amsmath,epsfig,amssymb,verbatim,amsopn,cite,multirow}
\usepackage{amsmath}
\usepackage{bm}
\usepackage{algorithm}
\usepackage{algorithmic}
\usepackage{subcaption}

\usepackage{color}

\usepackage{cite}
\usepackage{url}

\usepackage{array} 
\usepackage[margin=14.5mm,top=18.1mm,bottom=26.2mm]{geometry}

\usepackage[nodisplayskipstretch]{setspace}
\newcommand{\qa}{{\bf a}}

\newcommand{\qh}{{\bf h}}

\newcommand{\qn}{{\bf n}}

\newcommand{\qp}{{\bf p}}
\newcommand{\qq}{{\bf q}}

\newcommand{\qs}{{\bf s}}

\newcommand{\qu}{{\bf u}}

\newcommand{\qw}{{\bf w}}
\newcommand{\qx}{{\bf x}}
\newcommand{\qy}{{\bf y}}

\newcommand{\qA}{{\bf A}}

\newcommand{\qH}{{\bf H}}

\newcommand{\qP}{{\bf P}}

\newcommand{\qW}{{\bf W}}

\newcommand{\E}{\mathbb{E}}

\newcommand{\norm}[1]{\left\lVert #1 \right\rVert}
\newcommand{\abs}[1]{\left\lvert #1 \right\rvert}

\newcommand{\Kdl}{K_{\mathrm{dl}}}
\newcommand{\Kul}{K_{\mathrm{ul}}}
\newcommand{\Ntx}{N_{\mathrm{tx}}}
\newcommand{\Nrx}{N_{\mathrm{rx}}}
\newcommand{\NE}{N_{\mathrm{E}}}
\newcommand{\Tx}{{\mathrm{t}}}
\newcommand{\Rx}{{\mathrm{r}}}
\newcommand{\SI}{\mathrm{SI}}
\newcommand{\Ex}{\mathbb{E}}
\newcommand{\ssen}{\qs_{\mathrm{sen}}}

\newcommand{\ul}{\mathrm{ul}}
\newcommand{\dl}{\mathrm{dl}}

\newcommand{\xjul}{x_{\ul,j}[\ell]}

\newcommand{\Ev}{{\mathrm{E}}}
\newcommand{\BS}{\mathrm{BS}}

\newcommand{\UEuj}{{\mathrm{UE}}_j^{\ul}}
\newcommand{\UEdk}{{\mathrm{UE}}_k^{\dl}}

\begin{document}

\title{Privacy-Aware ISAC for Full-Duplex Monostatic Systems Using Movable Antennas}

\author{\IEEEauthorblockN{Yasas Savinda$^\dagger$, Mohammadali Mohammadi$^*$, Himal A. Suraweera$^\dagger$, Henk Wymeersch$^{\S}$}
\IEEEauthorblockA{$^\dagger$Department of Electrical and Electronic Engineering, University of Peradeniya, Sri Lanka}
\IEEEauthorblockA{$^*$Centre for Wireless Innovation (CWI), Queen's University Belfast, UK}
\IEEEauthorblockA{$^\S$ Department of Electrical Engineering, Chalmers University of Technology, Gothenburg, Sweden}
Emails: \{e19356, himal\}@eng.pdn.ac.lk, m.mohammadi@qub.ac.uk,  henkw@chalmers.se}
\maketitle

\begin{abstract}
This work investigates sensing privacy in full-duplex (FD) monostatic integrated sensing and communication (ISAC) systems with movable antennas (MAs). The proposed approach jointly optimizes beamforming and antenna trajectories to create a deceptive dummy DD-bin response at a passive sensing eavesdropper (Eve), while satisfying a true-bin sensing-quality requirement at the base station (BS). The resulting problem is highly non-convex. {To address this, a stage-wise alternating local-search framework is developed to obtain suboptimal solutions. Within this framework, we maximize the worst-case margin between dummy and true delay-Doppler (DD)-bin detector-oriented SINR surrogates over a discretized uncertainty region for Eve, incorporating detector-aligned dummy-bin refinement and true-bin preservation.} Simulation results show that the proposed MA-enabled design suppresses Eve's true-target DD-bin selection and increases dummy-bin selection probability compared with benchmark schemes, while maintaining reliable BS sensing performance.
\end{abstract}



\vspace{-0.5em}
\section{Introduction}
Driven by recent advances in ISAC, dual-functional wireless systems have emerged as a promising technology for sixth-generation networks \cite{LiuJSACISAC,ISACSignalProcessing}. By sharing spectrum, hardware, and signaling resources, ISAC enables communication and sensing to be jointly realized, improving spectral efficiency, reducing deployment cost, and supporting a wide range of wireless applications \cite{LiuJSACISAC,ISACSignalProcessing}.

Most ISAC research focuses on communication–sensing integration, waveform design, and performance trade-offs \cite{LiuJSACISAC,ISACSignalProcessing}, while sensing leakage remains relatively underexplored despite passive receivers’ ability to infer sensitive information from ISAC signals \cite{ISACPrivacySurvey}. To mitigate this threat, movable antennas (MAs) reconfigure channels through position adjustment, providing spatial degrees of freedom for signal enhancement, interference suppression, and geometry adaptation \cite{MAOverview,MASecureComm,MAArchitecture}. Artificial-noise beamforming and ambiguity-function engineering have also emerged to degrade unauthorized sensing \cite{AFEngineering}.

Sensing privacy and secure ISAC have been studied from several perspectives \cite{ANBeamforming,AFEngineering,LocalizationPrivacy,RISPrivacy,RenSecureCFISAC}. These include joint transmit and artificial-noise beamforming against a passive bistatic Eve with unknown location \cite{ANBeamforming}, ambiguity-function engineering to generate ghost targets \cite{AFEngineering}, privacy-aware beamforming for MIMO-OFDM localization \cite{LocalizationPrivacy}, RIS-based sensing-direction perturbation \cite{RISPrivacy}, and secure cell-free ISAC against information and sensing Eves \cite{RenSecureCFISAC}.

Prior works \cite{ANBeamforming,AFEngineering,LocalizationPrivacy,RISPrivacy,RenSecureCFISAC} enhance sensing privacy via signal design, artificial noise, beamforming, or reconfigurable intelligent surface-based channel control. However, they often assume fixed antenna arrays, focus on communication or localization privacy rather than DD deception, or rely on strong Eve assumptions. Unlike fixed arrays, MAs introduce an additional spatial degree of freedom by enabling dynamic position reconfiguration \cite{MAOverview,MASecureComm,MAArchitecture}. Motivated by this, we study a privacy-aware FD monostatic ISAC system with MAs, where a BS performs joint sensing and communication in the presence of a passive sensing Eve. Our main contributions are summarized as follows:
\begin{itemize}
    \item We propose an MA-enabled ISAC framework that enhances sensing privacy by jointly optimizing beamforming and pulse-to-pulse antenna positions. The key idea is to induce dummy DD-bin responses at Eve while satisfying the BS true-bin sensing-quality requirement.
    \item {We formulate the design as a non-convex optimization problem based on the ambiguity functions at the Eve and the BS, and develop a stage-wise AO heuristic to obtain suboptimal solutions.} Each subproblem is solved through a four-stage procedure to progressively refine the design. Simulation results show that the proposed approach transforms a privacy-vulnerable baseline into a privacy-preserving system while maintaining sensing performance.
\end{itemize}

\textit{Notation:} Bold uppercase (lowercase) letters denote matrices (vectors); $(\cdot)^\mathrm{T}$ and $(\cdot)^\mathrm{H}$ denote the transpose and conjugate transpose, respectively; $[x]_+\triangleq \max\{x,0\}$; $\E\{\cdot\}$ denotes the statistical expectation.

\vspace{-0.5em}
\section{System Model}\label{sec:system}
\vspace{-0.3em}
We consider a privacy-aware MA-aided monostatic ISAC system, in which a sensing Eve is discouraged from selecting the true target DD bin and is instead misled toward a designed dummy DD bin, even when exploiting ISAC signals as signals of opportunity. Specifically, an FD BS employs two MA-enabled uniform planar arrays (UPAs) for transmission and reception. The BS simultaneously serves $\Kdl$ downlink (DL) and $\Kul$ uplink (UL) single-antenna users while sensing a single target in the presence of a single-antenna Eve~\cite{PengMAFDISAC, DingNearFieldMAISAC}. Let $\mathcal K_\dl \triangleq \{1,\ldots,\Kdl\}$ and $\mathcal K_\ul \triangleq \{1,\ldots,\Kul\}$ denote the index sets of DL and UL users. The $k$-th DL user ($k\in\mathcal K_\dl$) and the $j$-th UL user ($j\in\mathcal K_\ul$) are denoted by $\UEdk$ and $\UEuj$, respectively.

The transmit and receive MAs move within two-dimensional regions $\mathcal{C}_t$ and $\mathcal{C}_r$ in the $x$--$y$ plane. At pulse index $\ell\in\mathcal L\triangleq\{1,\ldots,L\}$, their position matrices are
$\qP_\Tx[\ell]=[\qp_{\Tx,1}[\ell],\ldots,\qp_{\Tx,\Ntx}[\ell]]\in\mathbb{R}^{3\times\Ntx}$ and
$\qP_\Rx[\ell]=[\qp_{\Rx,1}[\ell],\ldots,\qp_{\Rx,\Nrx}[\ell]]\in\mathbb{R}^{3\times\Nrx}$,
where $\qp_{\Tx,n}[\ell]=[x_{\Tx,n}[\ell],y_{\Tx,n}[\ell],0]^\mathrm{T}$ and $\qp_{\Rx,i}[\ell]=[x_{\Rx,i}[\ell],y_{\Rx,i}[\ell],0]^\mathrm{T}$. Define $\mathcal{P}\triangleq\{\qP_\Tx[\ell],\qP_\Rx[\ell]\}_{\ell\in\mathcal L}$. Let $\mathcal N_t\triangleq\{1,\ldots,\Ntx\}$, $\mathcal N_r\triangleq\{1,\ldots,\Nrx\}$, $\mathcal L^{-}\triangleq\{1,\ldots,L-1\}$, $\mathcal I_t\triangleq\{(n,m)\in\mathcal N_t\times\mathcal N_t:n<m\}$, and $\mathcal I_r\triangleq\{(i,j)\in\mathcal N_r\times\mathcal N_r:i<j\}$. The antenna locations within $\mathcal{C}_t$ and $\mathcal{C}_r$ must then satisfy 
\begin{subequations}\label{eq:constraint_set}
\begin{align}
& \qp_{\Tx,n}[\ell]\in\mathcal{C}_t,\quad
  \qp_{\Rx,i}[\ell]\in\mathcal{C}_r,
  \quad \forall n\in\mathcal N_t, i\in\mathcal N_r, \ell\in\mathcal L, \label{eq:C1_new}\\
& \|\qp_{\Tx,n}[\ell+1]-\qp_{\Tx,n}[\ell]\|_2\le \Delta_{\Tx},
  \quad \forall n\in\mathcal N_t,\; \ell\in\mathcal L^{-}, \label{eq:C21_new}\\
& \|\qp_{\Rx,i}[\ell+1]-\qp_{\Rx,i}[\ell]\|_2\le \Delta_{\Rx},
  \quad \forall i\in\mathcal N_r,\; \ell\in\mathcal L^{-}, \label{eq:C22_new}\\
& \|\qp_{\Tx,n}[\ell]-\qp_{\Tx,m}[\ell]\|_2\ge d_{\Tx}^{\min},
  \quad \forall (n,m)\in\mathcal I_t,\; \ell\in\mathcal L, \label{eq:C23_new}\\
& \|\qp_{\Rx,i}[\ell]-\qp_{\Rx,j}[\ell]\|_2\ge d_{\Rx}^{\min},
  \quad \forall (i,j)\in\mathcal I_r,\; \ell\in\mathcal L, \label{eq:C24_new}
\end{align}
\end{subequations}
where $\Delta_{\Tx}$ and $\Delta_{\Rx}$ denote the maximum allowable pulse-to-pulse displacements of the transmit and receive antennas, respectively, while $d_{\Tx}^{\min}$ and $d_{\Rx}^{\min}$ specify the corresponding minimum inter-antenna spacing constraints.

The target is located at $\qp_{\mathrm{tar}} = [x_{\mathrm{tar}}, y_{\mathrm{tar}}, z_{\mathrm{tar}}]^\mathrm{T}$, while the sensing Eve resides within an uncertainty region $\mathcal{R}_{\Ev}$. For robust design, this region is discretized into a finite set of candidate locations $\mathcal{P}_{\Ev}=\{\qp_{\Ev,1},\ldots,\qp_{\Ev,\NE}\}\subseteq\mathcal{R}_{\Ev}$, where $\qp_{\Ev,r}$ denotes the $r$-th candidate location. The transmit and receive antennas move within planar regions and undergo bounded pulse-to-pulse micro-dither motion over a coherent processing interval (CPI) of $L$ pulses.\footnote{We assume LoS-dominated geometric propagation and perfect synchronization within each CPI. The BS has perfect CSI for the legitimate and modeled residual-SI links and knows the target-induced DD hypothesis used by Eve, but only Eve’s uncertainty region $\mathcal{R}_{\Ev}$, not her exact location. Relaxing these assumptions is left for future work.}


\vspace{-0.2em}
\subsection{Channel Model}
Under line-of-sight (LoS) sensing, let $\varphi$ and $\vartheta$ denote the elevation and azimuth angles to the target. The steering vectors for the $n$-th transmit and $i$-th receive MA elements are
\vspace{-0.1em}
\begin{align}
\big[\mathbf{a}_\Tx(\vartheta,\varphi;\qP_\Tx[\ell])\big]_n
&=\exp\!\big(-j\frac{2\pi}{\lambda}\qu^\mathrm{T}(\vartheta,\varphi) \qp_{\Tx,n}[\ell]\big),\\
\big[\mathbf{a}_\Rx(\vartheta,\varphi;\qP_\Rx[\ell])\big]_i
&=\exp\!\big(-j\frac{2\pi}{\lambda}\qu^\mathrm{T}(\vartheta,\varphi) \qp_{\Rx,i}[\ell]\big),
\end{align}
where $\lambda$ is the carrier wavelength and $\qu(\vartheta,\varphi)\!=\![
\cos\varphi\cos\vartheta,
\cos\varphi\sin\vartheta,
\sin\varphi]^\mathrm{T}\!\!\in\!\mathbb{R}^3$. As antenna positions vary with $\ell$, steering responses become slow-time dependent, consistent with geometry-dependent MA channel models.

The communication channels for $\UEdk$ and $\UEuj$ can be respectively modeled as functions of the transmit and receive MA position vectors, i.e.,
\begin{align*}
    \qh_{\dl,k}(\qP_\Tx[\ell]) &\!=\! 
    \alpha^{\dl}_{k} 
    \Big[e^{j\frac{2\pi}{\lambda} \Vert \qp_{\Tx,1}[\ell]-\qq^{\dl}_{k}\Vert },\ldots,
    e^{j\frac{2\pi}{\lambda} \Vert \qp_{\Tx,\Ntx}[\ell]-\qq^{\dl}_{k}\Vert }\Big]
    \\
    \qh_{\ul,j}(\qP_\Rx[\ell]) &\!=\! 
    \alpha^{\ul}_{j} 
    \Big[e^{j\frac{2\pi}{\lambda} \Vert \qp_{\Rx,1}[\ell]-\qq^{\ul}_{j}\Vert },\ldots,
    e^{j\frac{2\pi}{\lambda} \Vert \qp_{\Rx,\Nrx}[\ell]-\qq^{\ul}_{j}\Vert }\Big],
\end{align*}
where $\alpha^{\dl}_{k}$ and $\alpha^{\ul}_{j}$ denote the corresponding path-loss coefficients, while $\qq^{\dl}_{k}$ and $\qq^{\ul}_{j}$ represent the coordinates of $\UEdk$ and $\UEuj$, respectively.

SI is suppressed via three-stage cancellation: passive antenna isolation, RF analog cancellation, and baseband digital cancellation. Their combined effect is modeled by $0 < \rho_{\SI} \ll 1$, the SI suppression coefficient. Let $\qH_{\SI}\big(\qP_\Tx[\ell], \qP_\Rx[\ell]\big) \in \mathbb{C}^{\Nrx \times \Ntx}$ denote the residual SI channel matrix, whose $(i,n)$-th element is $\big[\qH_{\SI}\big(\qP_\Tx[\ell], \qP_\Rx[\ell]\big)\big]_{(i,n)}
= \rho_{\SI} \, h^{\SI}_{i,n}[\ell]$, where $h^{\SI}_{i,n}[\ell]$ is defined according to the uniform spherical wave model as
$h^{\SI}_{i,n}[\ell]
=
\exp\!\big(
-j\frac{2\pi}{\lambda}
\big\Vert\qp_{\Tx,n}[\ell]-\qp_{\Rx,i}[\ell]\big\Vert
\big)$~\cite{DingNearFieldMAISAC}.


\vspace{-0.6em}
\subsection{Signal Model}
The DL ISAC signal, used for simultaneous sensing and multi-user DL communication via the $\Ntx$-element MA array at time index $\ell$, can be expressed as
\vspace{-0.1em}
\begin{equation}\label{eq:tx}
\qx[\ell]
=\sum\nolimits_{k\in \mathcal K_\dl}\qw_k s_k[\ell]+\ssen[\ell],\quad \ell=1,\ldots, L,
\end{equation}
where $\qw_k \in \mathbb{C}^{\Ntx \times 1}$ denotes the DL beamforming vector for $\UEdk$; $s_k[\ell]$ is the corresponding DL symbol satisfying $\Ex\{|s_k[\ell]|^2\} = 1$; and $\ssen[\ell] \in \mathbb{C}^{\Ntx \times 1}$ with $P_{\mathrm{sen}}=\Ex\{\|\ssen[\ell]\|^2\}$ represents the sensing waveform, independent of $s_k[\ell]$. We assume that $\ssen[\ell]=c[\ell]\ssen$, where $c[\ell]$ is a known unit-modulus probing sequence and $\ssen$ is a fixed spatial sensing vector normalized to $P_{\mathrm{sen}}$. We set $\mathbf v_{\mathrm{sen}}=\ssen/\|\ssen\|$, so that $s_{\mathrm{ref}}[\ell]=\mathbf v_{\mathrm{sen}}^{H}\ssen[\ell]$. Let $\qW = [\qw_1, \dots, \qw_{K_{\mathrm{dl}}}] \in \mathbb{C}^{\Ntx \times \Kdl}$ denote the DL beamforming matrix. Accordingly, the received signal at $\UEdk$ is given by,
\vspace{-0.1em}
\begin{equation}\label{eq:dl_rx}
y_k^{\dl}[\ell]
\!=\!
\qh_{\dl,k}^{H}(\qP_\Tx[\ell])\qx[\ell]\! + \!
\sum\nolimits_{j\in \mathcal K_\ul} \sqrt{p_j} h_{jk}\xjul\! +\! n_k^{\dl}[\ell],
\end{equation}
where $n_k^{\dl}[\ell]\sim\mathcal{CN}(0,\sigma_{n}^{2})$ denotes the additive white Gaussian noise (AWGN) at $\UEdk$, $p_j$ is the transmit power of $\UEuj$, $h_{jk}$ is the channel between the $\UEuj$ and $\UEdk$, and $\xjul$ represents the UL signal from $\UEuj$, which satisfies $\Ex\{|\xjul|^2\} = 1$.


By substituting \eqref{eq:tx} into \eqref{eq:dl_rx}, the DL SINR of $\UEdk$ at pulse index $\ell$ is given by
\vspace{-0.5em}
\begin{equation}\label{eq:dl_sinr}
\mathrm{SINR}_{k}^{\dl}\!\big(\qW,\qP_\Tx[\ell]\big)
=
\frac{\Vert\qh_{\dl,k}^{H}(\qP_\Tx[\ell])\qw_k\Vert^{2}}
{I_{k}^{\dl}\!\big(\qW,\qP_\Tx[\ell]\big)},
\end{equation}
where $I_{k}^{\dl}\!\big(\qW,\qP_\Tx[\ell]\big)=
\sum\nolimits_{m\in\mathcal K_\dl,\,m\neq k}\Vert\qh_{\dl,k}^{H}(\qP_\Tx[\ell])\qw_m\Vert^{2} + \sum\nolimits_{j\in\mathcal K_\ul} p_j\vert h_{jk}\vert^2
+\Vert\qh_{\dl,k}^{H}(\qP_\Tx[\ell])\ssen[\ell]\Vert^{2}
+\sigma_{n}^{2}.$


The FD BS simultaneously receives the DL communication signal, the target-reflected signal, and UL transmissions from multiple users. Accordingly, the received signal at the BS can be expressed as~\cite{DingNearFieldMAISAC}
\vspace{-0.1em}
\begin{align}
\mathbf{y}[\ell]
&=  \sum\nolimits_{j\in \mathcal K_\ul}\sqrt{p_j}\qh_{\ul,j}(\qP_\Rx[\ell])\xjul
\nonumber\\
&+\alpha e^{j2\pi f_D \ell T_r}
\qA\big(\vartheta,\varphi;\qP_\Tx[\ell],\qP_\Rx[\ell]\big)
\qx[\ell-\ell_\tau] \nonumber\\
&\quad +\qH_{\SI}\big(\qP_\Tx[\ell],\qP_\Rx[\ell]\big)\qx[\ell]+\qn[\ell],
\vspace{0.2em}
\end{align}
where $\qA\big(\vartheta,\varphi;\qP_\Tx[\ell],\qP_\Rx[\ell]\big)\triangleq\qa_\Rx(\vartheta,\varphi;\qP_\Rx[\ell])\qa_\Tx^\mathrm{H}(\vartheta,\varphi;\qP_\Tx[\ell])$, $f_D$ denotes the Doppler frequency, and $\ell_\tau$ is a discrete delay-bin (matched-filter lag) index in the adopted DD-grid abstraction. $T_r$ determines the slow-time Doppler sampling. The coefficient $\alpha$ denotes the round-trip target coefficient, and $\qn[\ell]\sim\mathcal{CN}(\mathbf{0},\sigma_n^2\mathbf{I}_{\Nrx})$ is the BS receiver noise.
A passive sensing Eve, located at $\qp_{\Ev,r}\in\mathcal{R}_{\Ev}$, observes
\begin{equation}\label{eq:eve}
\begin{aligned}
z_\Ev[\ell]
&= 
{ e^{j2\pi f_D \ell T_r}}
h_{\mathrm{tar,E}}(\qp_{\Ev,r})
\qa_\Tx^\mathrm{H}\big(\vartheta,\varphi;\qP_\Tx[\ell]\big)\qx[\ell-\ell_\tau] \\
&\quad + 
\sum\nolimits_{j\in \mathcal K_\ul} \sqrt{p_j} h_{j\Ev}(\qp_{\Ev,r})\xjul +\eta_\Ev[\ell],
\vspace{0.2em}
\end{aligned}
\end{equation}
where $h_{\mathrm{tar,E}}(\qp_{\Ev,r})$ is the Eve-location-dependent channel between the target and Eve, $h_{j\Ev}(\qp_{\Ev,r})$ denotes the Eve-location-dependent channel between $\UEuj$ and sensing Eve, and $\eta_\Ev[\ell]\sim\mathcal{CN}(0,\sigma_n^2)$ is the AWGN at the sensing Eve. This passive observation model follows unauthorized sensing frameworks where Eve acts as a passive radar using reflected ISAC signals \cite{ANBeamforming,AFEngineering}. Eq.~\eqref{eq:eve} shows that Eve’s received phase depends on both the target–Eve geometry and MA positions.

For tractability, Eve uses a nominal target DD hypothesis, with location uncertainty captured by $h_{\mathrm{tar,E}}(\qp_{\Ev,r})$; spatial variations in the bistatic DD offset are neglected, and the direct BS--Eve component is assumed cancelled before DD processing.

\vspace{-0.1em}
\section{DD Statistics: Target Hypotheses and Guarded DD-bin SINR Surrogate}
Let $(\ell_{\tau,0},f_0)$ and $(\tilde{\ell}_{\tau},\tilde{f}_D)$ denote the true-target and designed dummy DD bins, respectively. Following standard matched-filter DD processing \cite{DuPCSOFDMISAC}, Eve's statistic at candidate location $\qp_{\Ev,r}\in\mathcal{P}_{\Ev}$ under hypothesis $(\ell_{\tau},f)$ is
\vspace{-0.1em}
\begin{equation}
\mathcal{A}_\Ev(\ell_{\tau},f;\qp_{\Ev,r})
\!=\!
\sum\nolimits_{\ell\in\mathcal L}
\!\!z_\Ev[\ell]\,e^{-j2\pi f\ell T_r}\,s_{\mathrm{ref}}^*[\ell-\ell_{\tau}],
\end{equation}
where $s_{\mathrm{ref}}[\ell]\triangleq \mathbf v_{\mathrm{sen}}^{H}\ssen[\ell]$, with $\mathbf v_{\mathrm{sen}}\in\mathbb{C}^{\Ntx\times 1}$ a fixed unit-norm sensing reference vector. Thus, $s_{\mathrm{ref}}[\ell]\in\mathbb{C}$ is the scalar reference sequence used by the BS and Eve matched filters for DD processing.

The corresponding BS matched-filter DD statistic is
\vspace{-0.1em}
\begin{equation}
\mathcal{A}_{\BS}(\ell_{\tau},f)
=
\sum\nolimits_{\ell\in\mathcal L}
\bar{y}_{\BS}[\ell]\,e^{-j2\pi f\ell T_r}\,s_{\mathrm{ref}}^*[\ell-\ell_{\tau}],
\end{equation}
where $\bar{y}_{\BS}[\ell]\triangleq \qu_{\BS}^{H}[\ell]\qy[\ell]$
denotes the scalar BS observation after receive combining, with $\qu_{\BS}[\ell]
=
\frac{\qa_{\Rx}(\vartheta,\varphi;\qP_\Rx[\ell])}
{\norm{\qa_{\Rx}(\vartheta,\varphi;\qP_\Rx[\ell])}}$,
where a target-direction matched receive combiner is adopted for simplicity and coherent accumulation of the true-target echo.

The BS and Eve DD power maps are given by the squared magnitudes of their matched-filter statistics, as in standard radar/ISAC processing \cite{DuPCSOFDMISAC}, i.e.,
\begin{subequations}\label{eq:powermap}
\begin{align}
D_{\BS}(\ell_{\tau},f)
&\triangleq \abs{\mathcal{A}_{\BS}(\ell_{\tau},f)}^2,\\
D_\Ev(\ell_{\tau},f;\qp_{\Ev,r})
&\triangleq \abs{\mathcal{A}_{\Ev}(\ell_{\tau},f;\qp_{\Ev,r})}^2.
\end{align}
\end{subequations}
Then, based on~\eqref{eq:powermap}, the true-bin and dummy-bin energies are defined as follows:
\vspace{-0.4em}
\begin{align}
S_{\BS}^{\mathrm{true}}
&\triangleq \E\!\left\{D_{\BS}(\ell_{\tau,0},f_0)\right\},\\
S_\Ev^{\mathrm{true}}(\qp_{\Ev,r})
&\triangleq \E\!\left\{D_\Ev(\ell_{\tau,0},f_0;\qp_{\Ev,r})\right\},\\
S_\Ev^{\mathrm{dum}}(\qp_{\Ev,r})
&\triangleq \E\big\{D_\Ev(\tilde{\ell}_{\tau},\tilde{f}_D;\qp_{\Ev,r})\big\}.
\end{align}

For fixed $\qp_{\Ev,r}$, these quantities represent average DD-bin energy, not an average over Eve locations. Optimization uses noise-free DD responses, with $\sigma_n^2$ added separately as the guarded-surrogate noise floor, whereas noisy observations are generated for empirical detector evaluation.

Due to the nonlinear dependence of detection probability on DD statistics \cite{RenSecureCFISAC,BehdadMultistatic}, we use a detector-oriented guarded DD-bin SINR surrogate. Let $\Omega$ denote the discrete DD grid, with the following index bins
\begin{equation}
(\ell_{\tau}^{\mathrm{true}},f^{\mathrm{true}})=(\ell_{\tau,0},f_0),
\quad
(\ell_{\tau}^{\mathrm{dum}},f^{\mathrm{dum}})=(\tilde{\ell}_{\tau},\tilde{f}_D).
\end{equation}

For each bin $\kappa\in\{\mathrm{true},\mathrm{dum}\}$, we partition $\Omega$ into three regions: a guard region $\mathcal{G}_{\kappa}\subset\Omega$ centered at $(\ell_{\tau}^{\kappa},f^{\kappa})$, a sidelobe ring $\mathcal{L}_{\kappa}\subset\Omega$, and a remainder set $\mathcal{Q}_{\kappa} \triangleq \Omega\setminus(\mathcal{G}_{\kappa}\cup\mathcal{L}_{\kappa})$. The guard region includes the bin and nearest DD neighbors and is excluded from interference computation, while the ring is a thin band capturing sidelobe leakage; its depth equals the number of delay/Doppler bins between its outer and inner boundaries.

By invoking~\eqref{eq:powermap}, for each candidate Eve location $\qp_{\Ev,r}\in\mathcal{P}_\Ev$, the desired signal term is defined as
\begin{align}\label{eq:sig}
S_\Ev^{\kappa}(\qp_{\Ev,r})
\triangleq
\E\!\left\{D_\Ev(\ell_{\tau}^{\kappa},f^{\kappa};\qp_{\Ev,r})\right\}, \kappa\in\{\mathrm{true},\mathrm{dum}\}.
\end{align}
Moreover, the local and global interference terms, 
$I_{\Ev,\mathrm{loc}}^{\kappa}(\qp_{\Ev,r})$ and 
$I_{\Ev,\mathrm{glob}}^{\kappa}(\qp_{\Ev,r})$, 
are defined as the average off-bin energies within the sidelobe ring and the remainder region, respectively:
\begin{align}
I_{\Ev,\mathrm{loc}}^{\kappa}(\qp_{\Ev,r})
&\!\triangleq\!
\frac{1}{|\mathcal{L}_{\kappa}|}
\sum\nolimits_{(\ell_{\tau},f)\in\mathcal{L}_{\kappa}}
\E\!\left\{D_\Ev(\ell_{\tau},f;\qp_{\Ev,r})\right\},\\
I_{\Ev,\mathrm{glob}}^{\kappa}(\qp_{\Ev,r})
&\!\triangleq\!
\frac{1}{|\mathcal{Q}_{\kappa}|}
\sum\nolimits_{(\ell_{\tau},f)\in\mathcal{Q}_{\kappa}}
\!\!\E\!\left\{D_\Ev(\ell_{\tau},f;\qp_{\Ev,r})\right\}.
\end{align}
Here, $I_{\Ev,\mathrm{loc}}^{\kappa}(\qp_{\Ev,r})$ captures localized sidelobe leakage, whereas $I_{\Ev,\mathrm{glob}}^{\kappa}(\qp_{\Ev,r})$ represents the off-bin background over the DD grid, analogous to constant false-alarm rate (CFAR)-style guard/training-cell separation \cite{RadarSCNRJDL}.

We adopt a weighted, guarded local/global decomposition as a detector-oriented modeling choice. Using a weighting factor $\omega \in [0,1]$, the interference at the Eve is defined as:
\begin{align}\label{eq:int}
I_\Ev^{\kappa}(\qp_{\Ev,r})
=
\omega\,I_{\Ev,\mathrm{loc}}^{\kappa}(\qp_{\Ev,r})
+
(1-\omega)\,I_{\Ev,\mathrm{glob}}^{\kappa}(\qp_{\Ev,r}).
\end{align}
Here, $\omega$ controls the trade-off between local sidelobe leakage and the off-bin background. Larger $\omega$ emphasizes the local component, while smaller $\omega$ gives more weight to global interference. 

Using~\eqref{eq:sig} and~\eqref{eq:int}, Eve's detector-oriented guarded DD-bin SINR surrogate is
\vspace{-0.4em}
\begin{equation}
\mathrm{SINR}_\Ev^{\kappa}(\qp_{\Ev,r})
=
\frac{S_\Ev^{\kappa}(\qp_{\Ev,r})}
{I_\Ev^{\kappa}(\qp_{\Ev,r})+\sigma_n^2},
\quad \kappa\in\{\mathrm{true},\mathrm{dum}\},
\end{equation}
rather than a conventional pre-detection physical SINR.

At the BS, we enforce a true-bin sensing-quality requirement at $(\ell_{\tau,0},f_0)$. Define $S_{\BS}^{\mathrm{true}}
=
\E\!\left\{D_{\BS}(\ell_{\tau,0},f_0)\right\}$, 
and the corresponding weighted interference as $I_{\BS}^{\mathrm{true}}
=
\omega\,I_{\BS,\mathrm{loc}}^{\mathrm{true}}
+
(1-\omega)\,I_{\BS,\mathrm{glob}}^{\mathrm{true}}$,
where the local and global terms are defined analogously from $D_{\BS}(\ell_{\tau},f)$. The corresponding guarded DD-bin SINR surrogate is
\vspace{-0.7em}
\begin{equation}
\mathrm{SINR}_{\BS}^{\mathrm{true}}
=
\frac{S_{\BS}^{\mathrm{true}}}
{I_{\BS}^{\mathrm{true}}+\sigma_{n}^2}.
\vspace{0.2em}
\end{equation}
\vspace{-0.6em}
\section{Privacy-Aware Design}
\vspace{-0.3em}
\subsection{Problem Formulation}
The design jointly optimizes pulse-indexed MA states and transmit beamforming to suppress Eve's true DD-bin response, promote a predefined dummy bin, and satisfy the BS true-bin sensing-quality requirement. The baseline problem is
\vspace{-0.2em}
\begin{subequations}\label{eq:P1}
\begin{alignat}{2}
\qP_1:&\!\!\!\max_{\{\qP_\Tx[\ell],\,\qP_\Rx[\ell]\},\,\qW}\;
&& \!\min_{\qp_{\Ev,r}\in\mathcal{P}_\Ev}
\!\!\!\big(
\mathrm{SINR}_{\Ev}^{\mathrm{dum}}(\qp_{\Ev,r})
-
\mathrm{SINR}_{\Ev}^{\mathrm{true}}(\qp_{\Ev,r})
\big) \nonumber\\ 
&\hspace{2.em}\text{s.t.} 
&            &
\hspace{-1.8em}\mathrm{SINR}_{\BS}^{\mathrm{true}} \ge \Gamma_{\BS}^{\mathrm{sinr}}, \label{eq:P1:ct1}\\
&         &      &\hspace{-1.9em}\mathrm{SINR}_k^{\dl}\!\big(\qW,\qP_\Tx[\ell]\big)\!\ge\! \gamma_k^{\dl},
  \; \forall k\!\in\!\mathcal K_{\dl}, \ell\!\in\!\mathcal L, \label{eq:P1:ct2}\\
&         & 
& \hspace{-1.8em}\mathbb{E}\!\left\{\|\qx[\ell]\|^2\right\}\le P_{\max},
  \quad \forall \ell\in\mathcal L, \label{eq:P1:ct3}\\
&&&\hspace{-1.8em}\eqref{eq:C1_new}\text{--}\eqref{eq:C24_new}.\label{eq:P1:ct4}
\end{alignat}
\end{subequations}
where~\eqref{eq:P1:ct1}--\eqref{eq:P1:ct3} impose the BS guarded DD-bin sensing-surrogate, DL quality-of-service (QoS), and transmit-power constraints, respectively.

Although ($\qP_1$) promotes privacy through Eve's worst-case dummy-to-true SINR margin, DD detection also depends on threshold crossing and competing off-bin peaks. We therefore introduce detector-aligned metrics for the final design.

Let $\mathcal{G}_{\mathrm{true}}\triangleq\mathcal{G}_{\kappa}\big|_{\kappa=\mathrm{true}}$ and $\mathcal{G}_{\mathrm{dum}}\triangleq\mathcal{G}_{\kappa}\big|_{\kappa=\mathrm{dum}}$ denote the guard regions of the true and dummy bins, respectively. The Eve-side off-bin set used for dummy detection is defined as
\begin{equation}
\Omega_{\mathrm{off}}^{\mathrm{dum}}
\triangleq
\Omega\setminus
\big(\mathcal{G}_{\mathrm{true}}\cup\mathcal{G}_{\mathrm{dum}}\big).
\label{eq:Omega_off_dum}
\end{equation}
Using the Eve DD power map in~\eqref{eq:powermap}, the strongest Eve off-bin competitor and the dummy-bin detection threshold are respectively defined as
\vspace{-0.2em}
\begin{subequations}
\begin{align}
\widetilde{S}_{\Ev,\mathrm{off}}(\qp_{\Ev,r})
\triangleq
\max_{(\ell_\tau,f)\in\Omega_{\mathrm{off}}^{\mathrm{dum}}}
\E\!\left\{D_{\Ev}(\ell_\tau,f;\qp_{\Ev,r})\right\},
\label{eq:Seoff_E}\\
\eta_{\Ev}^{\mathrm{dum}}(\qp_{\Ev,r})
\triangleq
\bar{D}_{\Ev,\mathrm{off}}(\qp_{\Ev,r})
+
\beta_{\mathrm{th}}\,\sigma_{\Ev,\mathrm{off}}(\qp_{\Ev,r}),
\label{eq:etaE_dum}    
\end{align}
\end{subequations}
where $\bar{D}_{\Ev,\mathrm{off}}$ and $\sigma_{\Ev,\mathrm{off}}$ are the mean and standard deviation over $\Omega_{\mathrm{off}}^{\mathrm{dum}}$, with $\beta_{\mathrm{th}}>0$. Eve selects the strongest threshold-exceeding DD bin; false alarm refers to selecting the designed dummy bin instead of the true bin. The Eve-grid averaged metrics are
\vspace{-0.5em}
\begin{align}
\bar{\Delta}_{\Ev}^{\mathrm{dom}}
&\triangleq
\frac{1}{\NE}\sum\nolimits_{r=1}^{\NE}
\frac{
S_{\Ev}^{\mathrm{dum}}(\qp_{\Ev,r})
-
S_{\Ev}^{\mathrm{true}}(\qp_{\Ev,r})
}{
S_{\Ev}^{\mathrm{dum}}(\qp_{\Ev,r})
+
S_{\Ev}^{\mathrm{true}}(\qp_{\Ev,r})
+\epsilon
},
\label{eq:avgDeltaE_dom}\\
\bar{\Delta}_{\Ev}^{\mathrm{thr}}
&\triangleq
\frac{1}{\NE}\sum\nolimits_{r=1}^{\NE}
\frac{
S_{\Ev}^{\mathrm{dum}}(\qp_{\Ev,r})
-
\eta_{\Ev}^{\mathrm{dum}}(\qp_{\Ev,r})
}{
\eta_{\Ev}^{\mathrm{dum}}(\qp_{\Ev,r})+\epsilon
},
\label{eq:avgDeltaE_thr}\\
\bar{\Delta}_{\Ev}^{\mathrm{pk}}
&\triangleq
\frac{1}{\NE}\sum\nolimits_{r=1}^{\NE}
\frac{
S_{\Ev}^{\mathrm{dum}}(\qp_{\Ev,r})
-
\widetilde{S}_{\Ev,\mathrm{off}}(\qp_{\Ev,r})
}{
S_{\Ev}^{\mathrm{dum}}(\qp_{\Ev,r})
+
\widetilde{S}_{\Ev,\mathrm{off}}(\qp_{\Ev,r})
+\epsilon
}.
\label{eq:avgDeltaE_pk}
\end{align}
Here, $\epsilon>0$ ensures numerical stability, $\bar{\Delta}_{\Ev}^{\mathrm{dom}}$ promotes dummy-bin dominance over the true bin, $\bar{\Delta}_{\Ev}^{\mathrm{thr}}$ enforces dummy-bin threshold crossing, and $\bar{\Delta}_{\Ev}^{\mathrm{pk}}$ promotes dominance over the strongest off-bin competitor. For BS true-bin sensing-quality control, define the strongest off-bin competitor as
\vspace{-0.1em}
\begin{equation}\label{eq:bs_off_peak}
\widetilde{S}_{\BS}^{\mathrm{off}}
\triangleq
\max_{(\ell_{\tau},f)\in\Omega\setminus\mathcal{G}_{\mathrm{true}}}
\E\!\left\{D_{\BS}(\ell_{\tau},f)\right\}.
\end{equation}
Following peak-sidelobe analysis in ambiguity-function radar \cite{PeerYangPSLAF}, define
\vspace{-0.9em}
\begin{equation}
\Xi_{\BS}
\triangleq
\frac{
S_{\BS}^{\mathrm{true}}-\widetilde{S}_{\BS}^{\mathrm{off}}
}{
S_{\BS}^{\mathrm{true}}+\widetilde{S}_{\BS}^{\mathrm{off}}+\epsilon
},
\quad
\Pi_{\BS}
\triangleq
\frac{
\widetilde{S}_{\BS}^{\mathrm{off}}
}{
S_{\BS}^{\mathrm{true}}+\epsilon
}.
\label{eq:XiPi_BS}
\end{equation}
The metric $\Xi_{\BS}$ rewards BS true-bin dominance, while $\Pi_{\BS}$ penalizes off-bin peaks. To support the BS true-bin sensing-quality requirement, we define the normalized true-bin threshold excess as
\vspace{-0.8em}
\begin{equation}
\Theta_{\BS}
\triangleq
\frac{
S_{\BS}^{\mathrm{true}}-\eta_{\BS}^{\mathrm{true}}
}{
\eta_{\BS}^{\mathrm{true}}+\epsilon
},
\label{eq:Theta_BS}
\end{equation}
where $\eta_{\BS}^{\mathrm{true}}$ denotes the BS detection threshold based on off-bin background outside the true-bin guard region. Using these metrics, the detector-aware privacy design is formulated as
\vspace{-0.1em}
\begin{subequations}\label{eq:bs_clean_refined_problem}
\begin{alignat}{2}
\qP_2:&\!\!\!\max_{\{\qP_\Tx[\ell],\,\qP_\Rx[\ell]\},\,\qW,\,t}\;
&&t
+\lambda_{\Ev,\mathrm{dom}}\bar{\Delta}_{\Ev}^{\mathrm{dom}}
+\lambda_{\Ev,\mathrm{thr}}\bar{\Delta}_{\Ev}^{\mathrm{thr}}
\nonumber\\
&&&\quad
+\lambda_{\Ev,\mathrm{pk}}\bar{\Delta}_{\Ev}^{\mathrm{pk}}
+\lambda_{\BS,\mathrm{dom}}\Xi_{\BS}
\nonumber\\
&&&\quad
-\lambda_{\BS,\mathrm{peak}}\Pi_{\BS}
+\lambda_{\BS,\mathrm{thr}}\Theta_{\BS} \nonumber\\ 
&\hspace{2.em}\text{s.t.} 
&            &
\hspace{-2em}\mathrm{SINR}_{\Ev}^{\mathrm{dum}}(\qp_{\Ev,r})
\!-\!
\mathrm{SINR}_{\Ev}^{\mathrm{true}}(\qp_{\Ev,r})
\!\ge\! t,
\label{eq:P2:ct1}\\
&         &      &\hspace{-1.9em}\eqref{eq:P1:ct1}\text{--}\eqref{eq:P1:ct4}.
\label{eq:P2:ct2}
\end{alignat}
\end{subequations}
where all $\lambda$ parameters are non-negative weights, and $t$ represents the worst-case Eve dummy-versus-true SINR margin. The Eve-side terms encourage dummy-bin selection, while the BS-side terms promote true-bin dominance and threshold excess for sensing-quality control.

\vspace{-0.5em}
\subsection{Optimization Problem Solution}
\vspace{-0.5em}
Problem ($\qP_2$) is non-convex because $\mathbf W$ and $\mathcal P$ are coupled in the channels, steering vectors, and DD responses. We apply a four-stage AO heuristic comprising multi-start warm start (MWS), hard-constrained (HC), exact-dummy detector-aligned (EDD), and BS-clean preserve-Eve (BCP) refinements. Let $\Phi_s$ and $\mathcal F_s$ denote the stage-$s$ objective and feasible set.

\textbf{\textit{Sub-problem 1: $\mathbf{W}$-design with $\mathcal{P}$ fixed:}}
To reduce complexity, we set $\mathbf{W}=\mathbf{B}(\mathcal{P})\mathbf{C}$, where $\mathbf{C}\in\mathbb{C}^{N_b\times K_{\dl}}$ and $\mathbf{B}(\mathcal P)$ orthonormalizes a reduced beamspace comprising the target, uniform-spatial, and conjugate DL-channel directions, supplemented as needed to dimension $N_b$. The real and imaginary parts of $\mathbf C$ are stacked for the Powell search, yielding
\begin{equation}
\qP_{2-1}:\max_{\mathbf{C}} \;\; \Phi_s\!\big(\mathbf{B}(\mathcal{P})\mathbf{C},\mathcal{P}\big)
\quad
\text{s.t.}\quad
\big(\mathbf{B}(\mathcal{P})\mathbf{C},\mathcal{P}\big)\in\mathcal{F}_s.
\label{eq:ao_beam_block_reduced}
\end{equation}
This problem is solved by derivative-free local search with normalization and feasibility checks.

\textbf{Sub-problem 2: $\mathcal{P}$-design with $\mathbf{W}$ fixed:}
The geometry-update block is written as
\begin{equation}
\qP_{2-2}:~\max_{\mathcal{P}} \;\; \Phi_s(\mathbf{W},\mathcal{P})
\quad
\text{s.t.}\quad
(\mathbf{W},\mathcal{P})\in\mathcal{F}_s.
\label{eq:ao_position_block_generic}
\end{equation}
This block is solved by perturbation-based local search over feasible MA trajectories satisfying the region, spacing, and pulse-to-pulse movement constraints.

\subsubsection{Stage-Wise AO Framework}
The four stages alternate between the beamforming update in ($\qP_{2-1}$) and the position update in ($\qP_{2-2}$). To link the stage-wise objectives to the final target in~\eqref{eq:bs_clean_refined_problem}, define the worst-case Eve margin as
\begin{equation}
\Psi_\Ev(\mathbf{W},\mathcal{P})
\!\triangleq\!\!\!\!
\min_{\qp_{\Ev,r}\in\mathcal{P}_\Ev}
\big(
\mathrm{SINR}^{\mathrm{dum}}_{\Ev}(\qp_{\Ev,r})
\!-\!
\mathrm{SINR}^{\mathrm{true}}_{\Ev}(\qp_{\Ev,r})
\big).
\label{eq:PsiE_stage}
\end{equation}

The BS and DL violation terms are
$\mathcal{V}_{\BS}=\big[\Gamma_{\BS}^{\mathrm{sinr}}-\mathrm{SINR}_{\BS}^{\mathrm{true}}\big]_+$
and
$\mathcal{V}_{\mathrm{DL}}=\sum_{\ell\in\mathcal L}\sum_{k\in\mathcal K_{\mathrm{dl}}}
\big[\gamma_k^{\mathrm{dl}}-\mathrm{SINR}_k^{\mathrm{dl}}(\mathbf W,\qP_\Tx[\ell])\big]_+$.

\textbf{\textit{Phase MWS (Multi-Start Warm Start):}} 
This phase initializes the non-convex search by improving Eve's worst-case margin, rewarding BS sensing quality, and penalizing DL QoS violations:
\begin{subequations}
\begin{alignat}{2}
&\max_{\mathbf{W},\mathcal{P}}\;
&&\quad
\begin{aligned}[t]
\Phi_{\mathrm{MWS}}(\mathbf{W},\mathcal{P})
\triangleq\;&
\Psi_\Ev(\mathbf{W},\mathcal{P})
\\
&\hspace{-3.5em}+\lambda_{\mathrm{BS,MWS}}
\log\!\big(1+\mathrm{SINR}_{\BS}^{\mathrm{true}}\big)
-\mu_{\mathrm{DL}}\mathcal{V}_{\mathrm{DL}},
\end{aligned}
\nonumber\\
&\hspace{0.5em}\text{s.t.}
&&\quad
(\mathbf{W},\mathcal{P})\in\mathcal{F}_{\mathrm{MWS}},
\label{eq:stageMWS_problem}
\end{alignat}
\end{subequations}
where $\mathcal{F}_{\mathrm{MWS}}
\triangleq
\big\{
(\mathbf{W},\mathcal{P}) :
\|\mathbf{W}\|_F^2 \le P_{\max}-P_{\mathrm{sen}},
\eqref{eq:C1_new}--\eqref{eq:C24_new}
\big\}$. Here, $\lambda_{\mathrm{BS,MWS}}\ge0$ is the soft BS sensing-reward weight and $\mu_{\mathrm{DL}}\ge0$ is the DL QoS penalty weight. Problem~\eqref{eq:stageMWS_problem} is solved over multiple feasible initializations.

\textbf{\textit{Phase HC (Hard-Constrained AO Refinement):}} 
Starting from the warm-start solution, Phase HC enforces the BS sensing and DL QoS constraints as hard feasibility conditions. Then,
\begin{subequations}
\begin{alignat}{2}
&\max_{\mathbf{W},\mathcal{P}}\;
&&\quad \Phi_{\mathrm{HC}}(\mathbf{W},\mathcal{P})
\triangleq
\Psi_\Ev(\mathbf{W},\mathcal{P})
\nonumber\\
&\hspace{0.5em}\text{s.t.} 
&            &\quad 
(\mathbf{W},\mathcal{P})\in\mathcal{F}_{\mathrm{HC}},
\label{eq:stageHC_problem}
\end{alignat}
\end{subequations}
where $\mathcal{F}_{\mathrm{HC}}
\triangleq
\big\{
(\mathbf{W},\mathcal{P}) :
\|\mathbf{W}\|_F^2 \le P_{\max}-P_{\mathrm{sen}},
\mathcal{V}_{\BS}=0,
\mathcal{V}_{\mathrm{DL}}=0,
(1a)\text{--}(1e)
\big\}$. 
During HC and the subsequent hard-feasible stages, a candidate beamforming or MA-position update is accepted only if the BS sensing and all DL QoS requirements are satisfied; otherwise, the previous feasible iterate is retained.


\textbf{\textit{Phase EDD (Exact-Dummy Detector-Aligned Refinement):}} 
Starting from the hard-feasible solution, Phase EDD activates the Eve detector-aligned part of the final target in~\eqref{eq:bs_clean_refined_problem}. The corresponding stage problem is
\vspace{-0.1em}
\begin{subequations}
\begin{alignat}{2}
&\max_{\mathbf{W},\mathcal{P}}\;
&&\quad \Phi_{\mathrm{EDD}}(\mathbf{W},\mathcal{P})
\triangleq
\Psi_{\Ev}(\mathbf{W},\mathcal{P})
+\lambda_{\Ev,\mathrm{dom}}\bar{\Delta}_{\Ev}^{\mathrm{dom}}
\nonumber\\
&&&\quad
+\lambda_{\Ev,\mathrm{thr}}\bar{\Delta}_{\Ev}^{\mathrm{thr}}
+\lambda_{\Ev,\mathrm{pk}}\bar{\Delta}_{\Ev}^{\mathrm{pk}}
\nonumber\\
&\hspace{0.5em}\text{s.t.} 
&            &\quad 
(\mathbf{W},\mathcal{P})\in\mathcal{F}_{\mathrm{EDD}},
\label{eq:stageEDD_problem}
\end{alignat}
\end{subequations}
with $\mathcal{F}_{\mathrm{EDD}}\triangleq\mathcal{F}_{\mathrm{HC}}$. Using AO search, this phase sharpens the dummy DD response while preserving $\mathcal{F}_{\mathrm{HC}}$ feasibility; a $0.02$-weighted penalty discourages Eve's true-bin threshold crossing.

\textbf{\textit{Phase BCP (BS-Clean Preserve-Eve Refinement):}} 
This phase activates the BS-clean part of the final target in~\eqref{eq:bs_clean_refined_problem}. It improves the legitimate BS true-bin prominence via $\Xi_{\BS}$ while penalizing strong BS off-bin competitors through $\Pi_{\BS}$. With initialization $\mathbf{W}^{(\mathrm{BCP},0)}=\mathbf{W}^{(\mathrm{EDD})\star}$, $\mathcal{P}^{(\mathrm{BCP},0)}=\mathcal{P}^{(\mathrm{EDD})\star}$, and $i^{(\mathrm{BCP})}=0$, where $i^{(\mathrm{BCP})}$ is the Phase-BCP AO iteration counter, the AO framework is applied to
\vspace{-0.1em}
\begin{subequations}
\begin{alignat}{2}
&\max_{\mathbf{W},\mathcal{P}}\;
&&\quad
\begin{aligned}[t]
\Phi_{\mathrm{BCP}}(\mathbf{W},\mathcal{P})
\triangleq\;&
\lambda_{\mathrm{wc}}\Psi_{\Ev}(\mathbf{W},\mathcal{P})
+\lambda_{\BS,\mathrm{dom}}\Xi_{\BS}\\
&-\lambda_{\BS,\mathrm{peak}}\Pi_{\BS}
+\lambda_{\BS,\mathrm{thr}}\Theta_{\BS},
\end{aligned}
\nonumber\\
&\hspace{0.5em}\text{s.t.}
&            &\quad
(\mathbf{W},\mathcal{P})\in\mathcal{F}_{\mathrm{BCP}},
\label{eq:stageBCP_problem}
\end{alignat}
\end{subequations}
where $\mathcal{F}_{\mathrm{BCP}} \triangleq \mathcal{F}_{\mathrm{HC}}$, $\lambda_{\mathrm{wc}} \ge 0$ controls the retained Eve privacy margin, and $\Theta_{\BS}$ is the normalized BS true-bin threshold excess. BCP penalizes reductions below $0.85/0.85/0.80/0.80$ of the EDD reference margin, dominance, threshold-excess, and peak-window metrics, respectively. As both blocks use derivative-free local search, the multi-stage AO heuristic yields feasible suboptimal solutions without global-optimality guarantees.

The overall heuristic is summarized in \textbf{Algorithm~\ref{alg:compact_ma_isac}}. 
\begin{algorithm}[t]
\caption{Stage-Wise AO for Privacy-Aware MA-ISAC}
\label{alg:compact_ma_isac}
\begin{algorithmic}[1]
\STATE \textbf{Input:} $\mathcal{C}_{t}$, $\mathcal{C}_{r}$, $\mathcal{P}_{\Ev}$, $R$, {$\{I_s\}_{s\in\{\mathrm{MWS},\mathrm{HC},\mathrm{EDD},\mathrm{BCP}\}}$}, $P_{\max}$, $P_{\mathrm{sen}}$, $\Gamma_{\BS}^{\mathrm{sinr}}$, $\{\gamma_k^{\dl}\}_{k\in\mathcal K_{\dl}}$, and weighting parameters.

\STATE \textbf{Phase MWS: Multi-start warm start}
\STATE Generate feasible initial $\mathcal{P}$ and $\mathbf{W}$.
\STATE Solve \eqref{eq:stageMWS_problem} over $R$ initializations using \eqref{eq:ao_beam_block_reduced} and \eqref{eq:ao_position_block_generic}.
\STATE Retain the best $\mathbf{W}^{(\mathrm{MWS})\star}$, $\mathcal{P}^{(\mathrm{MWS})\star}$, and $\Psi_{\Ev}^{(\mathrm{MWS})\star}$.

\STATE \textbf{Phase HC: Hard-constrained AO refinement}
\STATE Initialize $\mathbf{W}^{(\mathrm{HC},0)}$, $\mathcal{P}^{(\mathrm{HC},0)}$ from MWS and set $i^{(\mathrm{HC})}=0$.
\REPEAT
    \STATE Solve \eqref{eq:stageHC_problem} using \eqref{eq:ao_beam_block_reduced} and \eqref{eq:ao_position_block_generic}.
    \STATE $i^{(\mathrm{HC})}=i^{(\mathrm{HC})}+1$.
\UNTIL{convergence or $i^{(\mathrm{HC})}\ge I_{\mathrm{HC}}$}

\STATE \textbf{Phase EDD: Exact-dummy detector-aligned refinement}
\STATE Initialize $\mathbf{W}^{(\mathrm{EDD},0)}$, $\mathcal{P}^{(\mathrm{EDD},0)}$ from HC and set $i^{(\mathrm{EDD})}=0$.
\REPEAT
    \STATE Solve \eqref{eq:stageEDD_problem} using \eqref{eq:ao_beam_block_reduced} and \eqref{eq:ao_position_block_generic}; set $i^{(\mathrm{EDD})}=i^{(\mathrm{EDD})}+1$.
\UNTIL{convergence or $i^{(\mathrm{EDD})}\ge I_{\mathrm{EDD}}$}

\STATE \textbf{Phase BCP: BS-clean preserve-Eve refinement}
\STATE Initialize $\mathbf{W}^{(\mathrm{BCP},0)}$, $\mathcal{P}^{(\mathrm{BCP},0)}$ from EDD, set $i^{(\mathrm{BCP})}=0$.
\REPEAT
    \STATE Solve \eqref{eq:stageBCP_problem} using \eqref{eq:ao_beam_block_reduced} and \eqref{eq:ao_position_block_generic}, while preserving the Phase-EDD Eve-side reference levels.
    \STATE $i^{(\mathrm{BCP})}=i^{(\mathrm{BCP})}+1$.
\UNTIL{convergence or $i^{(\mathrm{BCP})}\ge I_{\mathrm{BCP}}$}

\STATE \textbf{Output:} $\mathbf{W}^{\star}=\mathbf{W}^{(\mathrm{BCP})\star}$, $\{\qP_\Tx^{\star}[\ell],\qP_\Rx^{\star}[\ell]\}_{\ell=1}^{L}=\mathcal{P}^{(\mathrm{BCP})\star}$, and $t^{\star}=\Psi_{\Ev}(\mathbf{W}^{\star},\mathcal{P}^{\star})$.
\end{algorithmic}
\end{algorithm}

\textbf{\textit{Computational complexity:}} We quantify the derivative-free Powell beamforming cost by objective evaluations. Let $N_{\mathrm{Pow}}$ denote the maximum evaluations per beamforming update and $N_{\mathrm{sw}}$ the maximum MA-position sweeps per AO iteration. Each sweep tests at most $4L(N_{\mathrm{tx}}\!+\!N_{\mathrm{rx}})$ perturbations; hence, overall complexity is
$\mathcal{O}\!\left(I_{\mathrm{tot}}\!\left[N_{\mathrm{Pow}}\!+\!4N_{\mathrm{sw}}L(N_{\mathrm{tx}}\!+\!N_{\mathrm{rx}})\right]C_{\mathrm{eval}}\right)$,
where $I_{\mathrm{tot}}\!=\!RI_{\mathrm{MWS}}\!+\!I_{\mathrm{HC}}\!+\!I_{\mathrm{EDD}}\!+\!I_{\mathrm{BCP}}$. For fixed antenna and user dimensions, direct DD-grid evaluation gives $C_{\mathrm{eval}}\!=\!\mathcal{O}(N_{\mathrm{E}}|\Omega|L)$.


\setlength{\textfloatsep}{0.0cm}

\begin{table}[!t]
\centering
\caption{\small Main algorithmic and reproducibility settings.}
\label{tab:alg_settings}
\vspace{-0.7em}
\scriptsize
\renewcommand{\arraystretch}{0.78}
\setlength{\tabcolsep}{1.6pt}
\begin{tabular}{|>{\centering\arraybackslash}p{0.23\columnwidth}|
                >{\centering\arraybackslash}p{0.69\columnwidth}|}
\hline
\textbf{Setting} & \textbf{Value} \\
\hline
Initialization & Random starts $R=2$; global seed $42$\\
\hline
AO stage limits & MWS/HC/EDD/BCP iterations: $4/4/3/3$ \\
\hline
Beam update & $N_b=5$; Powell search; 60 iterations/update \\
\hline
Position update & $\pm x/\pm y$ perturbations; step $0.006$ m, halved to $0.001$ m \\
\hline
{Penalty settings} & {Internal BS/DL search penalties: $1000/500$; $\lambda_{\mathrm{BS,MWS}}=0.03$.} \\
\hline
{EDD weights} & {$(\lambda_{\Ev,\mathrm{dom}},\lambda_{\Ev,\mathrm{thr}},\lambda_{\Ev,\mathrm{pk}})=(0.03,0.01,0.03)$; true-bin penalty $0.02$} \\
\hline
{BCP weights} & {$(\lambda_{\mathrm{wc}},\lambda_{\BS,\mathrm{dom}},\lambda_{\BS,\mathrm{peak}})=(0.35,0.08,0.02)$; $\lambda_{\BS,\mathrm{thr}}=0.05$; preserve $0.85/0.85/0.80/0.80$} \\
\hline
DD detector settings & $\beta_{\mathrm{th}}=2$; guard $(1,1)$; local ring $(2,2)$; $\omega=0.7$ \\
\hline
{Evaluation settings} & {$4\times4$ Eve uncertainty grid with 50 independent noise realizations per grid point.} \\
\hline
\end{tabular}
\end{table}

\section{Numerical Results}
\vspace{-1em}
We consider an FD monostatic MA-ISAC BS with $N_{\mathrm{tx}}=N_{\mathrm{rx}}=6$, serving $K_{\mathrm{dl}}=K_{\mathrm{ul}}=2$ users. The carrier operates at $f_c=30$ GHz ($\lambda=0.01$ m), and each CPI contains $L=16$ pulses with repetition interval $T_r=10^{-4}$ s. Total transmit power is $P_{\max}=1$, with $P_{\mathrm{sen}}=0.2$ allocated to sensing. All receiver noise powers are $\sigma_n^2=10^{-3}$, and the residual SI factor is $\rho_{\mathrm{SI}}=0.002$. The reported detector-decision probabilities use $50$ independent noise realizations per Eve-grid point. The true target is at $\mathbf{p}_{\mathrm{tar}}=(8,2,0)$ m. The DL users are at $(4,7,0)$ m and $(5,-7,0)$ m, while the UL users are at $(-4.5,6,0)$ m and $(-5,-6,0)$ m. We use $(\alpha_1^{\dl},\alpha_2^{\dl})=(0.18,0.16)$, $(\alpha_1^{\ul},\alpha_2^{\ul})=(0.20,0.17)$, $p_j=5\times10^{-5}$ for each UL user (total UL power $10^{-4}$), and path-loss exponent $\nu=2$. Eve's region is discretized into a $4\times4$ grid over $x\in[12,16]$ m and $y\in[-4,4]$ m, yielding $\NE=16$ candidate locations. This grid defines the uncertainty set for the proof-of-concept robust design; denser and off-grid validation is left for future work. The delay and Doppler grids are $\{0,\ldots,6\}$ and $\{-1875,-1250,-625,0,625,1250,1875\}$ Hz, respectively. The true-target and dummy hypotheses correspond to DD bins $(\ell_{\tau,0},f_0)=(5,0~\mathrm{Hz})$ and $(\tilde{\ell}_{\tau},\tilde{f}_D)=(1,1250~\mathrm{Hz})$. For the aperture sweep, $A_t+A_r=0.08~\mathrm{m}^2$ and $\rho\triangleq A_r/A_t$, giving square-region areas $A_t=0.08/(1+\rho)$ and $A_r=0.08\rho/(1+\rho)$. We use $d_{\Tx}^{\min}=d_{\Rx}^{\min}=0.01$ m and $\Delta_{\Tx}=\Delta_{\Rx}=0.012$ m. Here, pulse-indexed MA positions denote switched or electronically reconfigurable antenna states during the CPI, rather than continuous mechanical motion between pulses.

The BS sensing and per-user DL SINR thresholds are $\Gamma_{\mathrm{BS}}^{\mathrm{sinr}}=4$ and $\gamma_k^{\mathrm{dl}}=0.10$, respectively. Detection uses $\beta_{\mathrm{th}}=2$, i.e., the mean off-bin response plus two standard deviations, excluding guards around the true and dummy bins. Table~\ref{tab:alg_settings} summarizes the fixed AO settings across all $\rho$. Stage limits cap alternating updates, while coordinate perturbation generates feasible MA candidates through small $(\pm x,\pm y)$ shifts.

To evaluate the proposed design, we consider three benchmarks with the same feasible random MA initialization and power-normalized channel-conjugate DL beamformer. \textbf{Benchmark~1} uses this initialization without further optimization. \textbf{Benchmark~2} jointly optimizes beamforming and MA states to enhance BS detection without Eve-privacy terms. \textbf{Benchmark~3} fixes the MA states and optimizes the beamformer for worst-case Eve DD deception. Benchmarks~2 and~3 impose the same BS sensing and DL QoS constraints. The proposed scheme jointly optimizes both blocks through multi-start, stage-wise privacy-aware AO. All schemes share system and evaluation settings.

\begin{figure}[t]
\centering
\begin{subfigure}{0.235\textwidth}
    \centering
    \includegraphics[width=\linewidth]{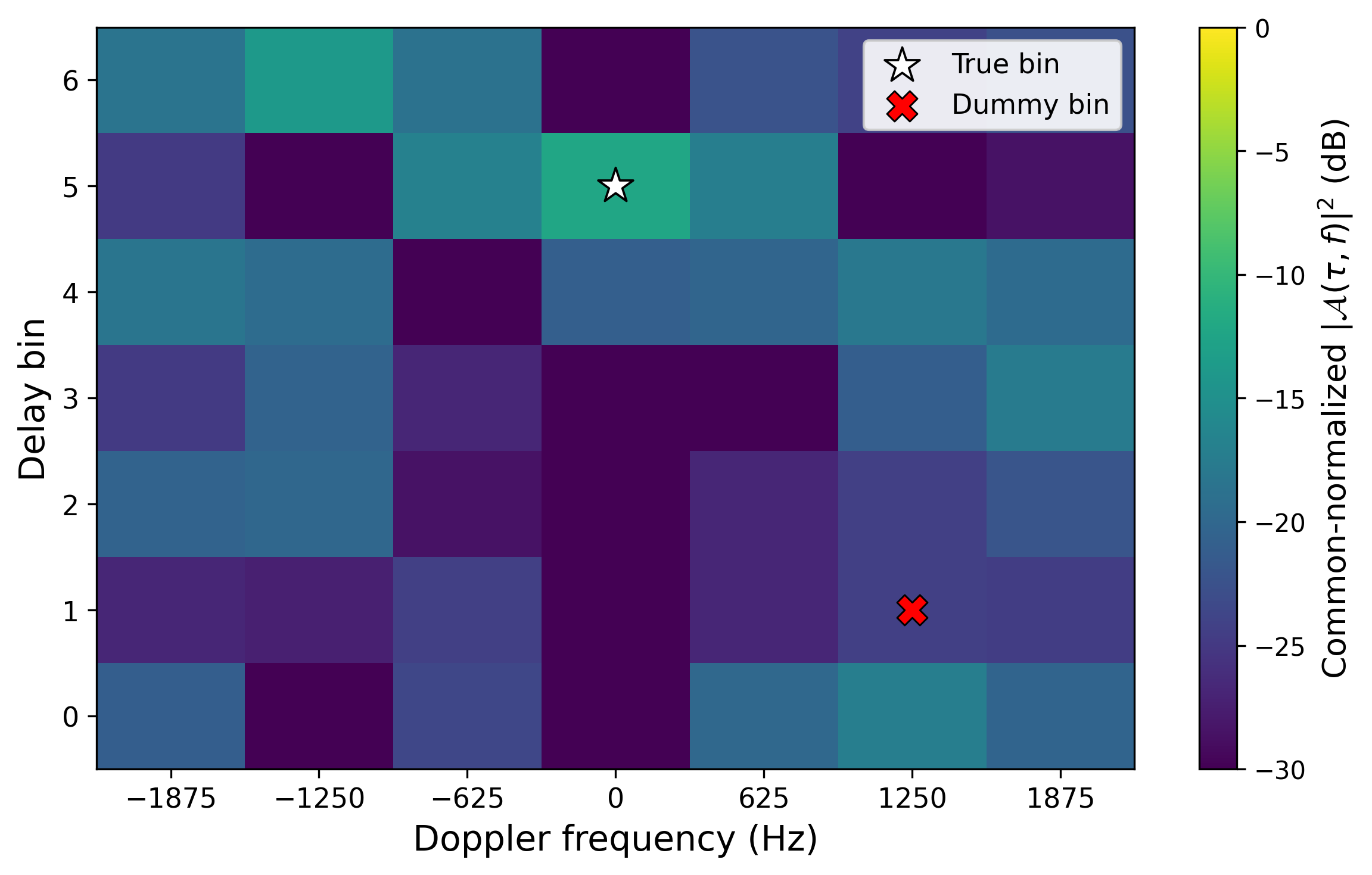}
    \caption{Benchmark 1}
    \label{fig:eve_dd_b1}
\end{subfigure}
\hfill
\begin{subfigure}{0.235\textwidth}
    \centering
    \includegraphics[width=\linewidth]{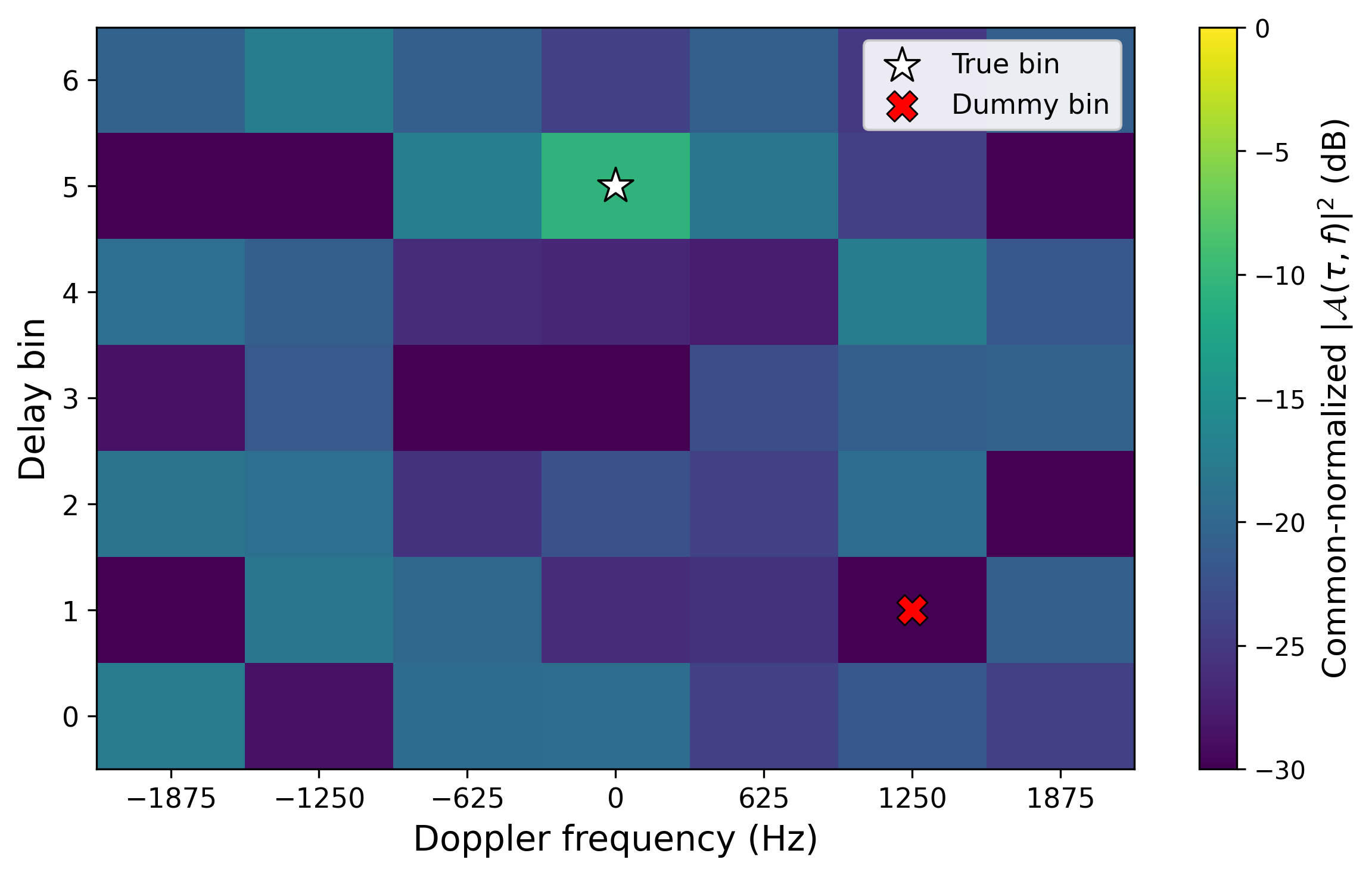}
    \caption{Benchmark 2}
    \label{fig:eve_dd_b2}
\end{subfigure}

\vspace{0.3em}

\begin{subfigure}{0.235\textwidth}
    \centering
    \includegraphics[width=\linewidth]{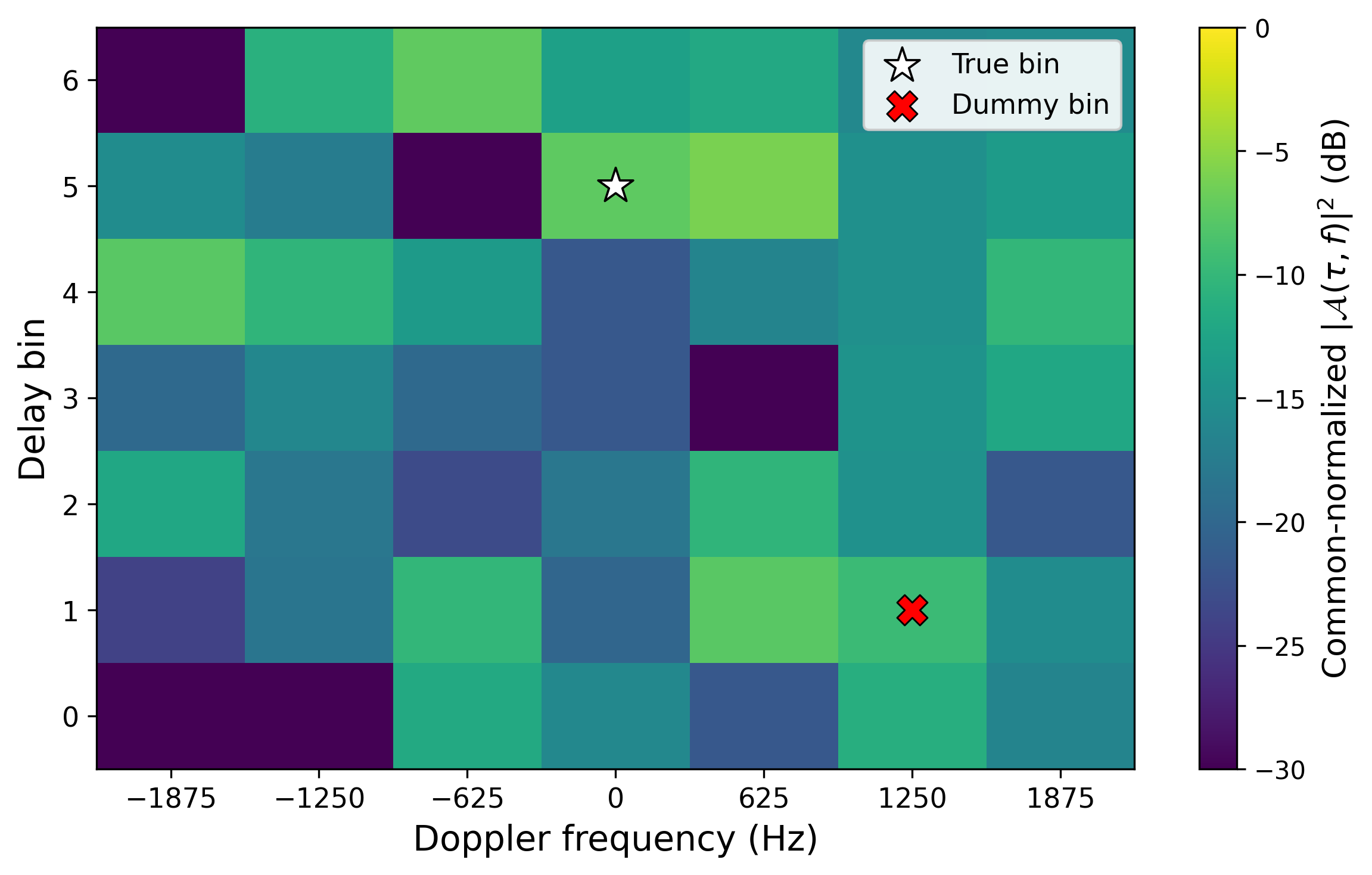}
    \caption{Benchmark 3}
    \label{fig:eve_dd_b3}
\end{subfigure}
\hfill
\begin{subfigure}{0.235\textwidth}
    \centering
    \includegraphics[width=\linewidth]{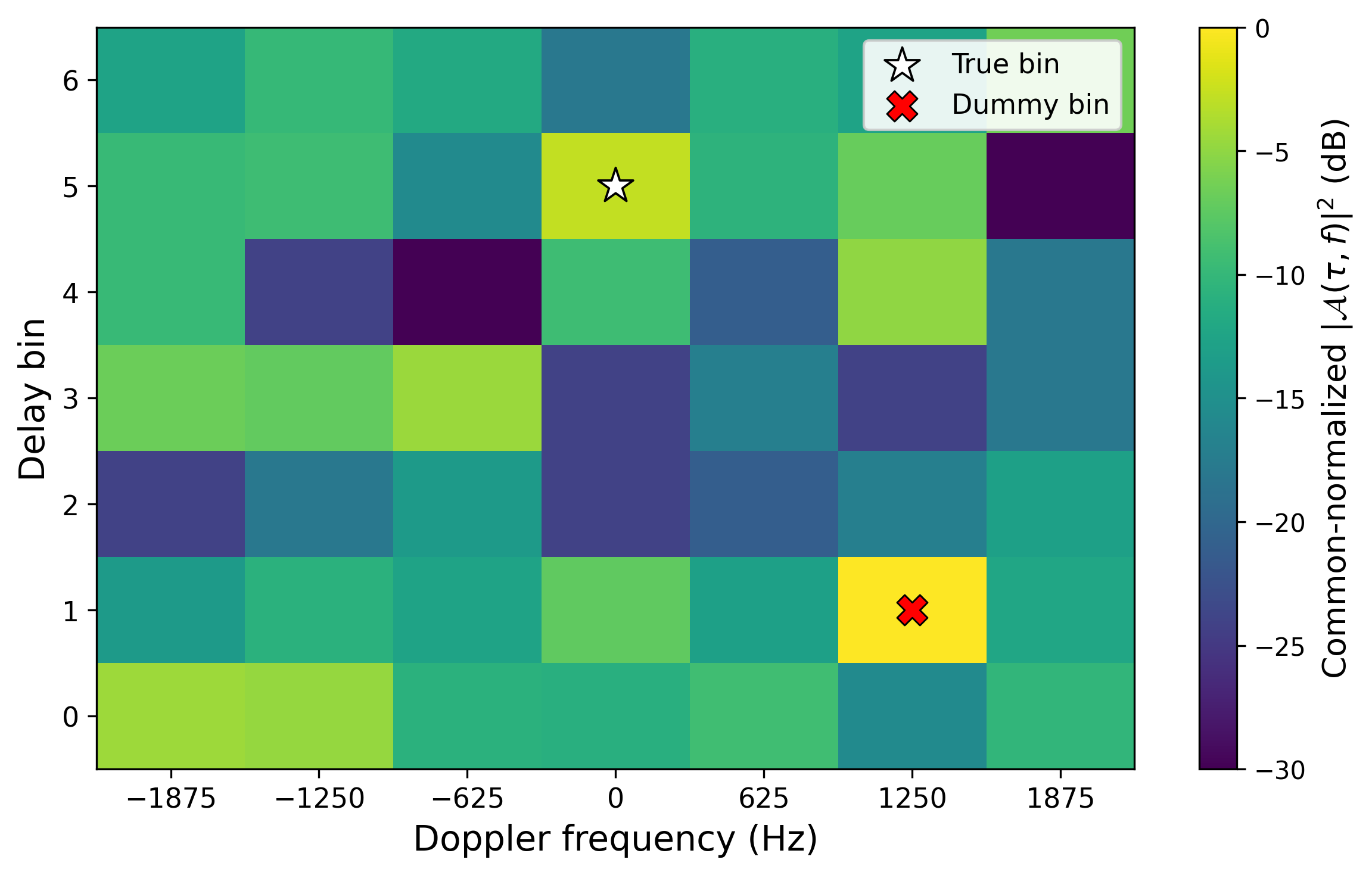}
    \caption{Proposed}
    \label{fig:eve_dd_proposed}
\end{subfigure}
\vspace{-0.7em}
\captionsetup[subfigure]{font=footnotesize}
\captionsetup[figure]{font=footnotesize}

\caption{\small Eve-side DD maps at $\rho=0.5$ for the benchmark schemes and proposed design. All maps are normalized by the common peak power across the four schemes and use the same $[-30,0]$-dB color scale.}
\label{fig:DDbin}
\vspace{-0.4em}
\end{figure}

\begin{figure}[t]
\vspace{-0.5em}
\centering

\begin{subfigure}[t]{0.48\columnwidth}
    \centering
    \includegraphics[trim=0 0cm 0cm 0cm,clip,width=\linewidth]{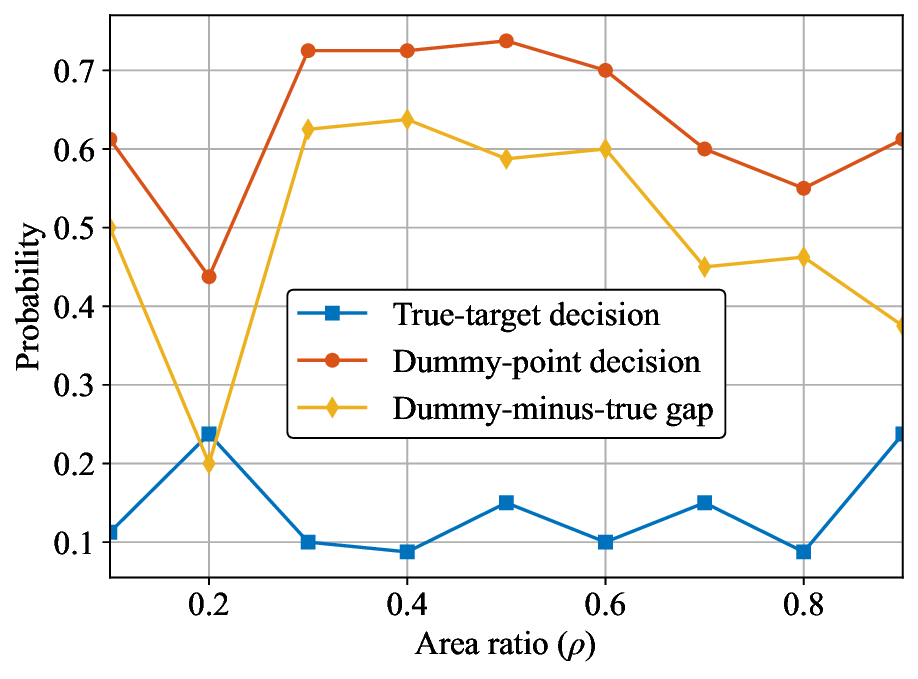}
    \caption{}
    \label{fig:eve_decision_gap}
\end{subfigure}
\hfill
\begin{subfigure}[t]{0.48\columnwidth}
    \centering
    \includegraphics[trim=0 0cm 0cm 0cm,clip,width=\linewidth]{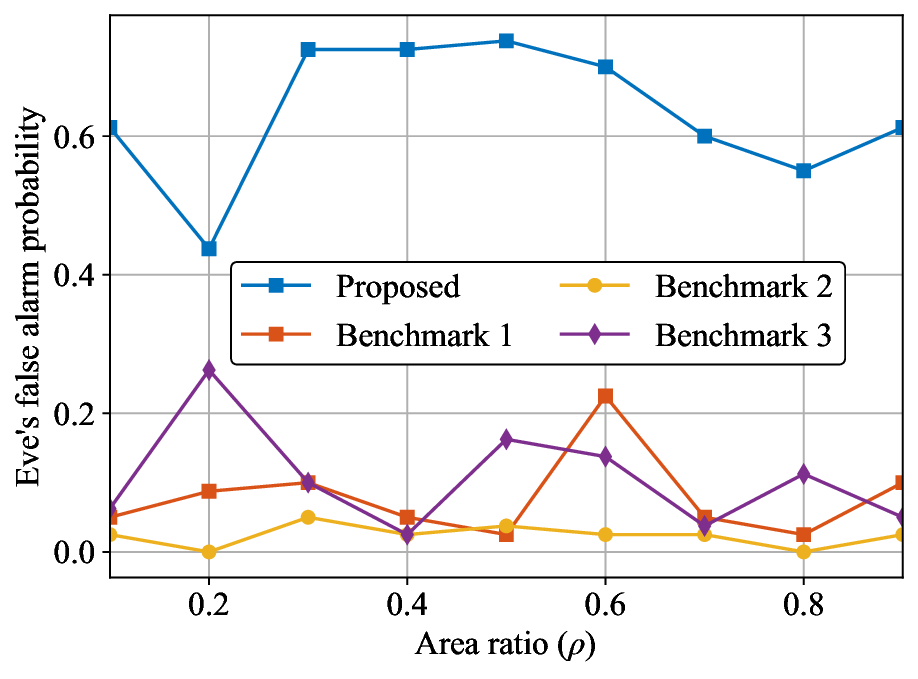}
    \caption{}
    \label{fig:eve_dummy_benchmark}
\end{subfigure}

\vspace{-0.6em}
\caption{\small Eve detection performance versus $\rho$: (a) proposed design; (b) Eve false-alarm probability, defined as selecting the dummy DD bin.}
\label{fig:ratio}
\vspace{0.2em}
\end{figure}

Figure~\ref{fig:DDbin} compares the DD maps at Eve under different designs. \textbf{Benchmark~1} and \textbf{Benchmark~2} leave a dominant response around the true target DD bin, while \textbf{Benchmark~3} only partially reduces this effect. In contrast, the \textbf{proposed} design shifts Eve's detector output toward the designed dummy DD bin, demonstrating the advantage of joint antenna location and beamforming optimization for DD-bin deception. This behavior indicates that the proposed scheme not only suppresses the true-target peak but also reshapes the DD landscape to create a more pronounced and misleading dummy peak, thereby increasing the likelihood of incorrect target identification at Eve.

Figure~\ref{fig:ratio}(a) shows Eve's detection performance versus the MA area ratio $\rho$. The proposed design maintains a higher dummy-bin than true-target decision probability, making Eve more likely to select the designed dummy DD bin. Figure~\ref{fig:ratio}(b) compares Eve's false-alarm probability, i.e., the probability of selecting the dummy bin as the target. The proposed design outperforms all benchmarks, particularly for $0.3<\rho<0.6$, showing that moderate receive-to-transmit MA aperture ratios strengthen DD-bin deception.

\vspace{-0.4em}
\section{Conclusion}
This paper studied DD-domain sensing privacy in FD monostatic MA-ISAC, where a passive Eve attempts to identify the true-target DD bin. Transmit beamforming and pulse-indexed MA states were jointly optimized to induce a deceptive dummy response at Eve while preserving BS true-bin sensing. The detector-aware formulation combines the worst-case dummy-to-true SINR margin, dummy-bin selection, and BS sensing control. A four-stage AO heuristic yielded feasible suboptimal solutions under sensing, communication, power, and MA constraints. Simulations showed the design suppresses Eve's true-bin selection and improves its dummy-bin selection over benchmarks while satisfying BS sensing and DL QoS requirements.

\vspace{0.2em}
\bibliographystyle{IEEEtran}
\bibliography{IEEEabrv,references}

@STRING{IEEE_J_SPL        = "{IEEE} Signal Process. Lett."}

@STRING{IEEE_J_SP         = "{IEEE} Trans. Signal Process."}

@STRING{IEEE_J_STSP = "{IEEE} J. Sel. Top. Sign. Proces"}

@STRING{IEEE_J_WCOM       = "{IEEE} Trans. Wireless Commun."}

@STRING{IEEE_J_WCOML       = "{IEEE} Wireless Commun. Lett."}

@STRING{IEEE_M_COM        = "{IEEE} Commun. Mag."}

@STRING{IEEE_M_WC         = "{IEEE} Wireless Commun. Mag."}

@STRING{IEEE_M_WC         = "{IEEE} Wireless Commun."}

@STRING{IEEE_J_JSAC       = "{IEEE} J. Sel. Areas Commun."}

@STRING{IEEE_J_WCOML      = "{IEEE} Wireless Commun. Lett."}

@STRING{IEEE_J_STSP       = "{IEEE} J. Sel. Topics Signal Process."}

@STRING{IEEE_J_ITM        = "{IEEE} Internet Things Mag."}

@article{RadarSCNRJDL,
  title={A single-dataset-based pre-processing joint domain localized algorithm for clutter-suppression in shipborne high-frequency surface-wave radar},
  author={Guo, Liang and Zhang, Xin and Yao, Di and Yang, Qiang and Bai, Yang and Deng, Weibo},
  journal={Sensors},
  volume={20},
  number={13},
  pages={3773},
  year={2020},
  publisher={MDPI}
}

@ARTICLE{RenSecureCFISAC,
  author={Ren, Zixiang and Xu, Jie and Qiu, Ling and Wing Kwan Ng, Derrick},
  journal=IEEE_J_JSAC, 
  title={Secure Cell-Free Integrated Sensing and Communication in the Presence of Information and Sensing Eavesdroppers}, 
  year={2024},
  volume={42},
  number={11},
  pages={3217-3231},
  ISSN={1558-0008},
  month={Nov},}

@ARTICLE{DuPCSOFDMISAC,
  author={Du, Zhen and Liu, Fan and Xiong, Yifeng and Han, Tony Xiao and Eldar, Yonina C. and Jin, Shi},
  journal=IEEE_J_SP, 
  title={Reshaping the ISAC Tradeoff Under {OFDM} Signaling: A Probabilistic Constellation Shaping Approach}, 
  year={2024},
  volume={72},
  number={},
  pages={4782-4797},
  ISSN={1941-0476},
  month={},}

@article{PengMAFDISAC,
  title={Movable Antenna Aided Full-Duplex {ISAC} System with Self-Interference Mitigation},
  author={Peng, Size and Xu, Yin and Yi, Guanli and Zhang, Cixiao and He, Dazhi and Zhang, Wenjun},
  journal={arXiv preprint},
  year={2025},
note={[Online]. Available: \url{https://arxiv.org/pdf/2505.14830}}
}

@ARTICLE{DingNearFieldMAISAC,
  author={Ding, Jingze and Zhou, Zijian and Shao, Xiaodan and Jiao, Bingli and Zhang, Rui},
  journal=IEEE_J_WCOM, 
  title={Movable Antenna-Aided Near-Field Integrated Sensing and Communication}, 
  year={2026},
  volume={25},
  number={},
  pages={493-508},
  month={},}

@article{RISPrivacy,
  title={Invisible Walls: Privacy-Preserving {ISAC} Empowered by Reconfigurable Intelligent Surfaces},
  author={He, Yinghui and Fan, Long and Xie, Lei and Niyato, Dusit and Yuen, Chau and Luo, Jun},
  journal={arXiv preprint},
  year={2026},
note={[Online]. Available: \url{https://arxiv.org/abs/2601.04488}}
}

@ARTICLE{LocalizationPrivacy,
  author={Zhang, Yuchen and Chen, Hui and Keskin, Musa Furkan and Pourafzal, Alireza and Zheng, Pinjun and Wymeersch, Henk and Al-Naffouri, Tareq Y.},
  journal=IEEE_J_WCOML, 
  title={Privacy Preservation in {MIMO-OFDM} Localization Systems: A Beamforming Approach}, 
  year={2025},
  volume={14},
  number={7},
  pages={1979-1983},
  ISSN={2162-2345},
  month={Jul.},}

@ARTICLE{AFEngineering,
  author={Han, Kawon and Meng, Kaitao and Masouros, Christos},
  journal=IEEE_J_WCOM, 
  title={Sensing-Secure {ISAC}: Ambiguity Function Engineering for Impairing Unauthorized Sensing}, 
  year={2026},
  volume={25},
  number={},
  pages={5386-5400},
  doi={10.1109/TWC.2025.3618121}}

@INPROCEEDINGS{ANBeamforming,
  author={Musallam, Ahmad and Li, Husheng},
  booktitle={Proc. IEEE ICC}, 
  title={Enhancing Sensing Privacy in ISAC Through Joint Signal and Artificial Noise Beamforming}, 
  year={2025},
  volume={},
  number={},
  pages={6019-6024},
  ISSN={1938-1883},
  month={June},}

@ARTICLE{MAArchitecture,
  author={Ning, Boyu and Yang, Songjie and Wu, Yafei and Wang, Peilan and Mei, Weidong and Yuen, Chau and Björnson, Emil},
  journal=IEEE_M_WC, 
  title={Movable Antenna-Enhanced Wireless Communications: General Architectures and Implementation Methods}, 
  year={2025},
  volume={32},
  number={5},
  pages={108-116},
  ISSN={1558-0687},
  month={Oct.},}

@ARTICLE{MASecureComm,
  author={Hu, Guojie and Wu, Qingqing and Xu, Kui and Si, Jiangbo and Al-Dhahir, Naofal},
  journal=IEEE_J_SPL  , 
  title={Secure Wireless Communication via Movable-Antenna Array}, 
  year={2024},
  volume={31},
  number={},
  pages={516-520},
  ISSN={1558-2361},
  month={},}

@ARTICLE{MAOverview,
  author={Zhu, Lipeng and Ma, Wenyan and Zhang, Rui},
  journal=IEEE_M_COM, 
  title={Movable Antennas for Wireless Communication: Opportunities and Challenges}, 
  year={2024},
  volume={62},
  number={6},
  pages={114-120},
  ISSN={1558-1896},
  month={June},}

@ARTICLE{ISACPrivacySurvey,
  author={Qu, Kaiqian and Ye, Jia and Li, Xuran and Guo, Shuaishuai},
  journal=IEEE_J_ITM, 
  title={Privacy and Security in Ubiquitous Integrated Sensing and Communication: Threats, Challenges and Future Directions}, 
  year={2024},
  volume={7},
  number={4},
  pages={52-58},
  ISSN={2576-3199},
  month={July},}

@ARTICLE{BehdadMultistatic,
  author={Behdad, Zinat and Demir, \"{O}zlem Tu\u{g}fe and Sung, Ki Won and Björnson, Emil and Cavdar, Cicek},
  journal=IEEE_J_WCOM , 
  title={Multi-Static Target Detection and Power Allocation for Integrated Sensing and Communication in Cell-Free Massive {MIMO}}, 
  year={2024},
  volume={23},
  number={9},
  pages={11580-11596},
  ISSN={1558-2248},
  month={Sep.},}

@ARTICLE{LiuJSACISAC,
  author={Liu, Fan and Cui, Yuanhao and Masouros, Christos and Xu, Jie and Han, Tony Xiao and Eldar, Yonina C. and Buzzi, Stefano},
  journal=IEEE_J_JSAC, 
  title={Integrated Sensing and Communications: Toward Dual-Functional Wireless Networks for {6G} and Beyond}, 
  year={2022},
  volume={40},
  number={6},
  pages={1728-1767},
  ISSN={1558-0008},
  month={Jun.},}

@ARTICLE{ISACSignalProcessing,
  author={Zhang, J. Andrew and Liu, Fan and Masouros, Christos and Heath, Robert W. and Feng, Zhiyong and Zheng, Le and Petropulu, Athina},
  journal=IEEE_J_STSP, 
  title={An Overview of Signal Processing Techniques for Joint Communication and Radar Sensing}, 
  year={2021},
  volume={15},
  number={6},
  pages={1295-1315},
  ISSN={1941-0484},
  month={Nov},}

@article{PeerYangPSLAF,
  author  = {Uri Pe'er and Ning Yang},
  title   = {Mathematical Analysis of the Peak Sidelobe Level of the Ambiguity Function for Random Phase Codes},
  journal = {IEEE Transactions on Aerospace and Electronic Systems},
  volume  = {58},
  number  = {5},
  pages   = {4669--4680},
  year    = {2022},
  doi     = {10.1109/TAES.2022.3164014}
}

\end{document}